\documentclass[article,twocolumn]{revtex4}

\usepackage{graphicx}
\usepackage{amsmath}
\usepackage{amssymb}
\usepackage{fancyhdr}
\usepackage{color}

\newcommand{\todoS}[1]{\texttt{\color{blue}#1}}

\newcommand{\ket}[1]{\left| #1 \right\rangle}

\newcommand{\braket}[2]{\left\langle #1 | #2 \right\rangle}

\newcommand{\be}{\begin{equation}}
\newcommand{\ee}{\end{equation}}
\newcommand{\bea}{\begin{eqnarray}}
\newcommand{\eea}{\end{eqnarray}}

\newcommand{\bean}{\begin{eqnarray*}}
\newcommand{\eean}{\end{eqnarray*}}

\renewcommand{\Pr}{{\rm Pr}}

\date{\empty}

\begin{document}
\title{Completing Quantum Mechanics with Quantized Hidden Variables}
\author{S.J.~van Enk}
\affiliation{
Department of Physics\\
University of Oregon, Eugene, OR 97403}
\date{\today}
\begin{abstract}
I explore the possibility that a quantum system S may be described {\em completely} by the combination of its standard quantum state $|\psi\rangle$ and a (hidden) quantum state
$|\phi\rangle$ (that lives in the same Hilbert space), such that the outcome of any standard projective measurement on the system S is determined once the two quantum states are specified. I construct an algorithm that retrieves the
standard quantum-mechanical probabilities, which depend only on $|\psi\rangle$, by assuming that the (hidden) quantum state $|\phi\rangle$ is drawn at random from some fixed probability distribution Pr(.) and by averaging over Pr(.). Contextuality and Bell nonlocality turn out to emerge automatically from this algorithm as soon as the dimension of the Hilbert space of S is larger than 2. If $|\phi\rangle$ is not completely random, subtle testable deviations from standard quantum mechanics may arise in sequential measurements on single systems.
\end{abstract}
\maketitle
\section{Introduction}
Standard textbook quantum mechanics does not 
provide any explanation of what causes one particular outcome to occur in a given measurement on a quantum system S prepared in some state $\ket{\psi}\in H_S$. 
The quest for such an explanation is tantamount to a search for hidden variables \cite{epr,bellepr,bell,chs,ks,peres,pbr}. 
Whereas hidden variables are usually thought of as being classical, here
I explore the possibility of using a hidden {\em quantum} variable, namely, a vector lying in the same Hilbert space $H_S$.
\todoS{(This same idea was first suggested by Bohm and Bub in 1966 \cite{bohm}, as Michael Hall pointed out to me. The algorithm I present is different from theirs, but the essence is the same.)}
I will describe an algorithm which assumes that the (hidden) vector is drawn at random  from some fixed probability distribution Pr(.), such that the standard quantum state and the hidden quantum state together determine the outcome of any measurement one may perform, and such that the correct quantum-mechanical probabilities of the possible measurement outcomes are obtained when averaging over the probability distribution Pr(.). 

I apply the algorithm, first to measurements on finite-dimensional systems (a single qubit, a single qutrit, and a pair of qubits), and subsequently to infinite-dimensional systems (a radioactively decaying atom (or nucleus), and the two-slit experiment).
\section{Algorithm}
Here is an algorithm that takes as input the standard quantum state $\ket{\psi}$ of some system S, a hidden quantum state $\ket{\phi_0}$, a probability distribution $\Pr(\ket{\phi})$ from which the hidden state has been drawn, and a standard projective measurement M on S. The algorithm's output consists of a single outcome for that measurement.  
Denote the Hilbert space dimension of system S by $D$, and assume that $D$ is finite (the infinite-dimensional case is discussed below). In that case, there are exactly $D$ outcomes for measurement M.
\begin{enumerate}
\item Calculate the Born-rule probabilities for each of the outcomes of M from the state $\ket{\psi}$
and sort them from largest to smallest. Write the sorted probabilities as $p_n=|\braket{n}{\psi}|^2$ for $n=1\ldots D$. (That is, the eigenstates of M are now denoted by $\{\ket{n}\}_{n=1\ldots D}$.)
\item Calculate corresponding Born-rule probabilities $q_n=|\braket{n}{\phi_0}|^2$ from the hidden state $\ket{\phi_0}$.
\item  Define $\pi_1$ such that the probability that $q_1$ is larger than $\pi_1$ equals $p_1$. In short-hand notation:
\[
{\rm Prob}(q_1>\pi_1)=p_1.
\]
That is,
define the subspace T of kets $\ket{\phi}$ such that
$|\braket{1}{\phi}|^2>\pi_1$ and adjust $\pi_1$ such that 
$\int_T \Pr(\ket{\phi}) d\ket{\phi}=p_1.$
\item 
Postulate that outcome 1 occurs iff
$q_1>\pi_1$.
\item Inductive rule: if outcome(s) $1\ldots n$ did not occur according to the algorithm,
then postulate that outcome $n+1$ occurs iff
\[q_{n+1}>\pi_{n+1},\] 
where $\pi_{n+1}$ is defined such that 
\[{\rm Prob'}(q_{n+1}>\pi_{n+1})=\frac{p_{n+1}}{1-p_1-p_2-...-p_n},\] 
where Prob' denotes a {\em conditional} probability for $q_{n+1}$, conditioned on the values of $q_1\ldots q_n$.
The right-hand side is the conditional probability that outcome $n+1$ occurs given that outcomes 1 through $n$ did not occur.
\end{enumerate}
This algorithm, by construction, will produce outcome $m$ with probability $p_m=|\braket{m}{\psi}|^2$ when one averages over Pr(.). Namely, by Rule 3, the probability with which outcome 1 is obtained equals $p_1$. And by Rule 5, once we have shown that outcomes 1 through $n$ occur with the correct probabilities $p_1\ldots p_n$, then outcome $n+1$ will occur with probability $p_{n+1}$. 

Two {\em pure} states are needed for having deterministic measurement outcomes. Whenever a system $S_1$ is entangled with some other system $S_2$ such that the joint state is pure but the state of $S_1$ is not, the algorithm is meant to be applied to the joint system.

The two quantum states $\ket{\psi}$ and $\ket{\phi_0}$ do not play a symmetric role in the algorithm, because (i) the algorithm is constructed so as to reproduce the probabilities according to the standard quantum state $\ket{\psi}$ (but not those of $\ket{\phi_0}$) and (ii) $\ket{\phi_0}$ is hidden and random. The standard quantum state is privileged in this sense.

In the next Section I will illustrate the algorithm by examples that were constructed so as not to need the inductive Rule 5: it is always the most likely outcome (according to $\ket{\psi}$) that occurs.
\section{The algorithm at work}
The algorithm does not specify the probability distribution Pr(.) over the states $\ket{\phi}$ (it works for any distribution). For a Hilbert space  $H^{(D)}$ of dimension $D<\infty$ one natural choice is the uniform distribution according to the unique unitarily invariant Haar measure. That is, the requirement on Pr(.) is that $\Pr(\ket{\phi})d\ket{\phi}=\Pr(U\ket{\phi})d(U\ket{\phi})$ for any unitary operator $U$ acting on $H^{(D)}$ and any $\ket{\phi}\in H^{(D)}$. In this Section the uniform distribution is the only choice considered.
\subsection{One qubit}
The rule provided by the algorithm turns out to be simple for a 2D Hilbert space if we use the unitarily invariant probability distribution Pr$(\ket{\phi}$). For any projective measurement there are just two outcomes, and we denote the larger of the two corresponding probabilities by $p_1=|\braket{1}{\psi}|^2$.
The condition on $q_1=|\braket{1}{\phi_0}|^2$ for the corresponding outcome to occur is 
\[
q_1>\pi_1:=1-p_1.
\]
An alternative way of writing the condition for picking outcome 1 rather than 2, is
$q_1+p_1>q_2+p_2$ \footnote{Here the set of states for which $q_1+p_1=p_2+q_2$ is ignored as that set is of measure zero. Analogous sets of measure zero for the higher-dimensional cases are ignored, too.}. This shows that the two quantum states do end up playing symmetric roles after all. As such, the two-dimensional case is similar to Spekkens' toy model \cite{spekkens}, in which an elementary physical system is modeled by assuming the underlying physical state is classical and contains two bits of information, but where one bit of information always remains unknown. This restricted state of knowledge then turns out to mimic surprisingly many features of a quantum state.
In the present model, on the other hand, there are two {\em quantum} bits of information, one of which is assumed unknown. This then reproduces all of the quantum mechanics of a qubit.
\subsection{Contextuality for a qutrit}
Let us move up one dimension and consider a qutrit.
The  most probable outcome of a  measurement on a qutrit in a given state $\ket{\psi}$ occurs, according to the algorithm,  when \footnote{For a D-dimensional Hilbert space $\pi_1=1-p_1^{1/(D-1)}$. This follows from the fact that  for a fixed ket $\ket{1}$ and for a ket $\ket{\phi}$ picked at random from the unitarily invariant probability distribution on a $D$-dimensional complex Hilbert space, Prob($|\braket{1}{\phi}|^2>c)=(1-c)^{D-1}$ for any constant $c$ s.t. $0\leq c\leq 1$.}
\[q_1>\pi_1:=1-\sqrt{p_1}.\]
This condition is not symmetric between $p_1$ and $q_1$. By way of example, consider a qutrit (viewed here as a spin-1 particle, with the eigenstates of $S_z$ chosen as basis states) 
in the state
\[
\ket{\psi}=\sqrt{0.26}\ket{m_z=-1}
+\sqrt{0.25}\ket{m_z=0}+\sqrt{0.49}\ket{m_z=+1},
\]
and assume that the hidden state is
\[
\ket{\phi}=\sqrt{0.3}\ket{m_z=-1}
+\sqrt{0.3}\ket{m_z=0}+\sqrt{0.4}\ket{m_z=+1}.
\]
A measurement of $S_z$ will yield the outcome $m_z=+1$ \footnote{because $p_1=0.49$ and $q_1=0.4>1-\sqrt{p_1}=0.3$.}.
On the other hand, if we measure a different observable, namely one with eigenstates
$\ket{\pm}:=(\ket{m_z=-1}\pm\ket{m_z=0})/\sqrt{2},\ket{m_z=1}$, then the outcome
will be $\ket{+}$ \footnote{because $p_1\approx 0.51$ and $q_1=0.6>1-\sqrt{p_1}\approx 0.29$.}, even though the outcome $\ket{m_z=+1}$ is still a possible outcome of this measurement, too. This feature is exactly the type of contextuality that appears in the Kochen-Specker theorem. That is, if this feature is explicitly excluded as a possibility for a hidden variable theory, then the theorem shows that a hidden-variable model thus restricted cannot reproduce all quantum-mechanical predictions.
\subsection{Bell nonlocality for two qubits}
When considering joint measurements on two qubits in an entangled pure state they are to be treated as a single measurement on a 4D system. The condition on $q_1$ for outcome 1 to occur is now
\[q_1>\pi_1:=1-
 \sqrt[\leftroot{-1}\uproot{2}\scriptstyle 3]{p_1}.
\]
Suppose we have two qubits, denoted by A and B, in the entangled state
\[
\ket{\Psi}_{A,B}=\sqrt{0.5}\ket{0}_A\ket{0}_B
+\sqrt{0.4}\ket{1}_A\ket{1}_B+\sqrt{0.1}\ket{0}_A\ket{1}_B,\]
and suppose that the hidden state is
a product state 
\[
\ket{\Phi}_{A,B}=\ket{\phi}_A\ket{\varphi}_B,
\]
with
\[
\ket{\phi}_A=\sqrt{0.8}\ket{0}_A+\sqrt{0.2}\ket{1}_A,
\]
and
\[
\ket{\varphi}_B=\sqrt{0.4}\ket{0}_B+\sqrt{0.6}\ket{1}_B.
\]
If we perform measurements on the pair of qubits in the standard basis (meaning that both A and B are measured in the basis $\{\ket{0},\ket{1}\}$) the outcome will be $\ket{0}_A\ket{0}_B$ \footnote{because $p_1=0.5$ and $q_1=0.32>1-p_1^{1/3}\approx 0.21$.}.
But when we perform a different measurement on A, namely, one in the basis $\ket{\pm}=(\ket{0}\pm\ket{1})/\sqrt{2}$, while B is still measured in the standard basis, the  joint measurement will have outcome 
$\ket{+}_A\ket{1}_B$ \footnote{because $p_1=0.45$ and $q_1=0.54>1-p_1^{1/3}\approx 0.23$.}, and so the measurement outcome on qubit B depends on what measurement is performed on A. This violates explicitly one of the assumptions that leads to Bell's inequalities.
The algorithm is thus explicitly nonlocal.
\todoS{(This very same observation was made about the Bohm-Bub model by Mattuck in 1981 \cite{mattuck}.)}
This nonlocality cannot be used for signaling as long as $\ket{\phi_0}$ is unknown.
On the other hand, if one knew the hidden state $\ket{\phi_0}$ one certainly could use that knowledge to achieve signaling. 
Hence, if Nature does not allow signaling, then a restriction must exist on how much knowledge of $\ket{\phi_0}$ is possible. I leave it as an open question to find exactly what upper limit on knowledge of the hidden state is set by a no-signaling constraint. 
\section{The algorithm in infinite dimensions}
The Haar measure does not exist in infinite-dimensional Hilbert spaces and we cannot pick a uniform probability distribution over states in such a space. Instead, a fairly natural alternative can be given for a separable Hilbert space (with a countable basis): given a system S and its Hamiltonian, 
we could assign a probability distribution over all possible states that depends explicitly on the energy, where, for example, we could assume uniform distributions (according to the corresponding Haar measures) on finite-dimensional degenerate subspaces of fixed energy. 
Alternatively, we could place an upper limit on the energy of the system and thereby define a finite-dimensional Hilbert subspace. Then we could use the Haar measure on that subspace to define the uniform distribution. The following examples are chosen so as to be largely independent of the choice of probability distribution for the hidden state. 

One sometimes uses infinite-dimensional Hilbert spaces that are not separable. For example, the theoretical description of a spontaneously broken symmetry, as in the Higgs mechanism or the ferromagnetic phase of iron, does require the concept of unitarily inequivalent Hilbert spaces. The algorithm for such a case needs yet another technical adjustment \footnote{For example, for the Heisenberg ferromagnet consisting of infinitely many spin-1/2 systems, we may parameterize the ground state manifold by two Euler angles, namely, by indicating the direction of each individual spin by those angles (and all spins in the ground state then point in that single direction). We could then postulate some distribution over the Euler angles. An excited state can be defined by flipping some finite number of spins of a given ground state. The hidden state can then be drawn from a distribution over ground states and the excited states thus defined.}, but
one still can define a probability measure over all such inequivalent Hilbert spaces. And thus the algorithm presented here could attribute a spontaneously broken symmetry to an asymmetric hidden quantum state.
\subsection{Phases of light and of Bose-Einstein condensates}
Consider an interference experiment in which two light beams with exactly $N$ photons each impinge on a 50/50 beamsplitter. Photodetectors (assumed perfect) placed at the two output ports of the beamsplitter
count photons.
If in one output port $N-n$ photons are counted then the other detector must  count
$N+n$ photons (with $-N\leq n\leq N$).
One may compare this experiment 
with a similar one with two light beams described by coherent states (``quasi-classical'' states \cite{cohen}), both with amplitude $|\alpha|=\sqrt{N}$ and a relative phase
difference $\vartheta$ determined by
$\sin\vartheta=\frac{n}{N}$. In the latter case one expects unequal counts of photons in a ratio  of approximately $(n+N)/(n-N)$, {\em because} the input beams have a well-defined phase difference. On the other hand, in the experiment with number states there is no such phase difference in the input state, but the measurement may very well produce unequal counts nonetheless (see \cite{molmer} for a nice illustration how such a phase difference builds up during the measurement). 

With the algorithm this particular outcome would be explained by  the hidden quantum state of two output light beams: for example,
if that state (just before counting photons, but after the beamsplitter) were
\[
\ket{\phi_0}=\sqrt{0.8}\ket{N-n}\ket{N+n}
+i\sqrt{0.1}\ket{0}\ket{0}-\sqrt{0.1}\ket{4}\ket{2},
\]
(note there is no reason the hidden state should contain exactly $2N$ photons)
then the explanation for the observed counting rates would be that the result was predetermined by the combination of the actual output state and the hidden state (especially its large first term). More generally, according to the algorithm, light beams may be said to contain a fixed number of photons {\em and} possess a well-defined phase difference (as revealed by an interference experiment), too.

A similar story may be told about the phase of Bose-Einstein condensates \cite{java}. Even if two independently prepared condensates have no well-defined phase difference, according to the standard quantum description, an interference experiment will still display crisp interference fringes \cite{andrews} as if a well-defined phase difference already existed. Again, according to the algorithm that measured phase difference is predetermined, thanks to the hidden state of the condensates.

\subsection{Radioactive decay \& spontaneous emission}
A radioactively decaying nucleus or atom has always been the prime example of an indeterministic system, for which the quantum laws of nature seem to introduce irreducible and objective chances:
The atom will decay at some point in time,
quantum mechanics provides the probability distribution for those decay times, but nothing in the atom or anywhere else in the universe determines what that point in time is going to be.

The algorithm does say something about when the atom will decay. Since the atom itself will not remain in a pure state, one could assume monitoring not just the mass (or energy) of the atom, but the surroundings of the atom and detect the decay product. For example, in the case of spontaneous decay by emission of a photon, the time of detection would be predetermined by the hidden state of the joint system consisting of the atom and the relevant modes of the electromagnetic field. The initial state of that joint system would be pure, with no photons present in the relevant modes and the atom in the excited state, and it would stay pure, as Mollow has shown in his analysis of this problem \cite{mollow}.
This pure state and the hidden state together determine when the atom decays.
\subsection{Two-slit experiment}
The mystery of the two-slit experiment lies in the combination of three propositions: (i) that single particles build up an interference pattern on a screen behind the two slits, (ii) that each particle must go through one slit only, but (iii) that with just a single slit open the interference pattern disappears.

Here is what the algorithm says about the two-slit experiment. Each particle will end up on a particular position on the screen, as determined by its wave function $\psi(\vec{r})$ and its hidden wave function $\phi(\vec{r})$; repeating the experiment will produce the standard interference pattern, because although $\psi(\vec{r})$ stays the same from one particle to the next, the hidden wave function varies randomly.
The algorithm also gives a definite answer (either ``slit 1'' or ``slit 2'') to the {\em counterfactual} question: ``If we had measured through which slit the particle went, what would we have found?''
But, just as on the standard quantum-mechanical account, an actual measurement will destroy the interference pattern (because the wave function does collapse upon measurement).
\section{Discussion}
Perhaps it is no surprise that one can concoct a hidden-variable model, necessarily nonlocal, contextual, and violating the Pusey-Barrett-Rudolph criterion of independence, such that the hidden variables conspire to reproduce the probabilistic results of quantum mechanics. What is more surprising is that there exists a very simple model achieving all this, in which a pair of quantum states---the standard state $\ket{\psi}$ and a hidden state $\ket{\phi_0}$---fully describes the physical state of a system S in that the pair determines the outcomes of all measurements on S (including incompatible ones). 

The extension of standard quantum mechanics through the addition of a hidden quantum state evades no-go theorems against particular types of hidden variables by violating at least one of the underlying assumptions. As regards the three main no-go theorems, those by Kochen-Specker, by Bell, and by Pusey, Barrett and Rudolph, the algorithm has these three consequences/properties:
(i) if the outcome of a measurement in a basis $\{\ket{a},\ket{b},\ket{c}\}$ is $\ket{a}$, the measurement outcome in the basis
$\{\ket{a},\ket{B},\ket{C}\}$ could be $\ket{B}$; (ii)
in a joint measurement on two systems A and B, the outcome of the measurement on A may depend on the measurement performed on system B; the algorithm is thus nonlocal; (iii) in a joint non-separable measurement on systems A and B, the hidden state may be entangled even if A and B were prepared independently. 

Bell inequality violations (as well as Hardy's paradox \cite{hardy} and, by an obvious extension to measurements on three systems, the GHZ paradox \cite{ghz}) are all explained by the non-locality property (ii).

A weak point of the algorithm as presented here is that it takes measurements as primitive, just as the standard account of quantum mechanics does, too. 
On the other hand, compared to the many-worlds interpretation, there is an economy of resources:
there are just two non-branching wave functions as opposed to one gigantic branching wave function.
Compared to Bohmian mechanics, which assign reality to both positions of corpuscles and the wave function, the algorithm is more symmetric: it assigns reality to two wave functions. 

There are further avenues for exploration. I did not discuss how the hidden state evolves in time and to what state it collapses after a measurement (nor what precisely constitutes a measurement).  One could assume it obeys the same Schr\"odinger equation as does $\ket{\psi}$ and that it
collapses to a random state upon measurement (and that random state can be said to determine to what eigenstate $\ket{\psi}$ collapses).
Alternatively, one could assume that the hidden quantum
state is not completely random and not completely hidden.
This could lead to subtle deviations from standard quantum-mechanical predictions, not concerning the probabilities of single measurement outcomes, but about sequences of measurements performed on a single system. For example, consider a sequence of measurements on a single spin-1/2 system: first we measure $S_z$, then $S_x$, then $S_z$ again, then $S_x$ again., etcetera. Standard QM predicts that the measurement outcomes are truly random and uncorrelated.
But if the hidden state $\ket{\phi_0}$ 
is actually not completely randomized after each measurement, it might be that the  $S_x$ measurement outcomes are pairwise correlated (either positively or negatively), and similarly for the $S_z$ measurement outcomes. \todoS{(This observation, too, was
made in a discussion of the Bohm-Bub model and an experiment was performed (in 1966!) to test this sort of prediction on photons (but not a single photon!) traveling through 3 polarizers \cite{pap}.)} 
In any case, a no-signaling 
requirement puts a restriction on how much information one may possess about $\ket{\phi_0}$.
\section*{Acknowledgements}
I thank Tommaso Calarco for enjoyable and useful discussions, and Michael Hall for pointing out the Bohm-Bub model to me.

\bibliography{symmetry}

\begin{thebibliography}{18}
\expandafter\ifx\csname natexlab\endcsname\relax\def\natexlab#1{#1}\fi
\expandafter\ifx\csname bibnamefont\endcsname\relax
  \def\bibnamefont#1{#1}\fi
\expandafter\ifx\csname bibfnamefont\endcsname\relax
  \def\bibfnamefont#1{#1}\fi
\expandafter\ifx\csname citenamefont\endcsname\relax
  \def\citenamefont#1{#1}\fi
\expandafter\ifx\csname url\endcsname\relax
  \def\url#1{\texttt{#1}}\fi
\expandafter\ifx\csname urlprefix\endcsname\relax\def\urlprefix{URL }\fi
\providecommand{\bibinfo}[2]{#2}
\providecommand{\eprint}[2][]{\url{#2}}

\bibitem[{\citenamefont{Einstein et~al.}(1935)\citenamefont{Einstein, Podolsky,
  and Rosen}}]{epr}
\bibinfo{author}{\bibfnamefont{A.}~\bibnamefont{Einstein}},
  \bibinfo{author}{\bibfnamefont{B.}~\bibnamefont{Podolsky}}, \bibnamefont{and}
  \bibinfo{author}{\bibfnamefont{N.}~\bibnamefont{Rosen}},
  \bibinfo{journal}{Phys. Rev.} \textbf{\bibinfo{volume}{47}},
  \bibinfo{pages}{777} (\bibinfo{year}{1935}).

\bibitem[{\citenamefont{Bell}(1964)}]{bellepr}
\bibinfo{author}{\bibfnamefont{J.~S.} \bibnamefont{Bell}},
  \bibinfo{journal}{Physics} \textbf{\bibinfo{volume}{1}}, \bibinfo{pages}{195}
  (\bibinfo{year}{1964}).

\bibitem[{\citenamefont{Bell}(1966)}]{bell}
\bibinfo{author}{\bibfnamefont{J.~S.} \bibnamefont{Bell}},
  \bibinfo{journal}{Rev. Mod. Phys.} \textbf{\bibinfo{volume}{38}},
  \bibinfo{pages}{447} (\bibinfo{year}{1966}).

\bibitem[{\citenamefont{Clauser et~al.}(1969)\citenamefont{Clauser, Horne,
  Shimony, and Holt}}]{chs}
\bibinfo{author}{\bibfnamefont{J.~F.} \bibnamefont{Clauser}},
  \bibinfo{author}{\bibfnamefont{M.~A.} \bibnamefont{Horne}},
  \bibinfo{author}{\bibfnamefont{A.}~\bibnamefont{Shimony}}, \bibnamefont{and}
  \bibinfo{author}{\bibfnamefont{R.~A.} \bibnamefont{Holt}},
  \bibinfo{journal}{Phys. Rev. Lett.} \textbf{\bibinfo{volume}{23}},
  \bibinfo{pages}{880} (\bibinfo{year}{1969}).

\bibitem[{\citenamefont{Kochen and Specker}(1967)}]{ks}
\bibinfo{author}{\bibfnamefont{S.}~\bibnamefont{Kochen}} \bibnamefont{and}
  \bibinfo{author}{\bibfnamefont{E.}~\bibnamefont{Specker}},
  \bibinfo{journal}{J. Math. Mech.} \textbf{\bibinfo{volume}{17}},
  \bibinfo{pages}{59} (\bibinfo{year}{1967}).

\bibitem[{\citenamefont{Peres}(1991)}]{peres}
\bibinfo{author}{\bibfnamefont{A.}~\bibnamefont{Peres}},
  \bibinfo{journal}{Journal of Physics A: Math. Gen.}
  \textbf{\bibinfo{volume}{24}}, \bibinfo{pages}{L175} (\bibinfo{year}{1991}).

\bibitem[{\citenamefont{Pusey et~al.}(2012)\citenamefont{Pusey, Barrett, and
  Rudolph}}]{pbr}
\bibinfo{author}{\bibfnamefont{M.~F.} \bibnamefont{Pusey}},
  \bibinfo{author}{\bibfnamefont{J.}~\bibnamefont{Barrett}}, \bibnamefont{and}
  \bibinfo{author}{\bibfnamefont{T.}~\bibnamefont{Rudolph}},
  \bibinfo{journal}{Nature Physics} \textbf{\bibinfo{volume}{8}},
  \bibinfo{pages}{475} (\bibinfo{year}{2012}).

\bibitem[{\citenamefont{Bohm and Bub}(1966)}]{bohm}
\bibinfo{author}{\bibfnamefont{D.}~\bibnamefont{Bohm}} \bibnamefont{and}
  \bibinfo{author}{\bibfnamefont{J.}~\bibnamefont{Bub}},
  \bibinfo{journal}{Reviews of Modern Physics} \textbf{\bibinfo{volume}{38}},
  \bibinfo{pages}{453} (\bibinfo{year}{1966}).

\bibitem[{\citenamefont{Spekkens}(2007)}]{spekkens}
\bibinfo{author}{\bibfnamefont{R.~W.} \bibnamefont{Spekkens}},
  \bibinfo{journal}{Phys. Rev. A} \textbf{\bibinfo{volume}{75}},
  \bibinfo{pages}{032110} (\bibinfo{year}{2007}).

\bibitem[{\citenamefont{Mattuck}(1981)}]{mattuck}
\bibinfo{author}{\bibfnamefont{R.~D.} \bibnamefont{Mattuck}},
  \bibinfo{journal}{Physics Letters A} \textbf{\bibinfo{volume}{81}},
  \bibinfo{pages}{331} (\bibinfo{year}{1981}).

\bibitem[{\citenamefont{Cohen-Tannoudji
  et~al.}(1997)\citenamefont{Cohen-Tannoudji, Dupont-Roc, and
  Grynberg}}]{cohen}
\bibinfo{author}{\bibfnamefont{C.}~\bibnamefont{Cohen-Tannoudji}},
  \bibinfo{author}{\bibfnamefont{J.}~\bibnamefont{Dupont-Roc}},
  \bibnamefont{and} \bibinfo{author}{\bibfnamefont{G.}~\bibnamefont{Grynberg}},
  \emph{\bibinfo{title}{Photons and Atoms-Introduction to Quantum
  Electrodynamics}}, vol.~\bibinfo{volume}{1} (\bibinfo{year}{1997}).

\bibitem[{\citenamefont{M{\o}lmer}(1997)}]{molmer}
\bibinfo{author}{\bibfnamefont{K.}~\bibnamefont{M{\o}lmer}},
  \bibinfo{journal}{Physical Review A} \textbf{\bibinfo{volume}{55}},
  \bibinfo{pages}{3195} (\bibinfo{year}{1997}).

\bibitem[{\citenamefont{Javanainen and Yoo}(1996)}]{java}
\bibinfo{author}{\bibfnamefont{J.}~\bibnamefont{Javanainen}} \bibnamefont{and}
  \bibinfo{author}{\bibfnamefont{S.~M.} \bibnamefont{Yoo}},
  \bibinfo{journal}{Phys. Rev. Lett.} \textbf{\bibinfo{volume}{76}},
  \bibinfo{pages}{161} (\bibinfo{year}{1996}).

\bibitem[{\citenamefont{Andrews et~al.}(1997)\citenamefont{Andrews, Townsend,
  Miesner, Durfee, Kurn, and Ketterle}}]{andrews}
\bibinfo{author}{\bibfnamefont{M.}~\bibnamefont{Andrews}},
  \bibinfo{author}{\bibfnamefont{C.}~\bibnamefont{Townsend}},
  \bibinfo{author}{\bibfnamefont{H.-J.} \bibnamefont{Miesner}},
  \bibinfo{author}{\bibfnamefont{D.}~\bibnamefont{Durfee}},
  \bibinfo{author}{\bibfnamefont{D.}~\bibnamefont{Kurn}}, \bibnamefont{and}
  \bibinfo{author}{\bibfnamefont{W.}~\bibnamefont{Ketterle}},
  \bibinfo{journal}{Science} \textbf{\bibinfo{volume}{275}},
  \bibinfo{pages}{637} (\bibinfo{year}{1997}).

\bibitem[{\citenamefont{Mollow}(1975)}]{mollow}
\bibinfo{author}{\bibfnamefont{B.}~\bibnamefont{Mollow}},
  \bibinfo{journal}{Physical Review A} \textbf{\bibinfo{volume}{12}},
  \bibinfo{pages}{1919} (\bibinfo{year}{1975}).

\bibitem[{\citenamefont{Hardy}(1992)}]{hardy}
\bibinfo{author}{\bibfnamefont{L.}~\bibnamefont{Hardy}},
  \bibinfo{journal}{Phys. Rev. Lett.} \textbf{\bibinfo{volume}{68}},
  \bibinfo{pages}{2981} (\bibinfo{year}{1992}).

\bibitem[{\citenamefont{Greenberger et~al.}(1990)\citenamefont{Greenberger,
  Horne, Shimony, and Zeilinger}}]{ghz}
\bibinfo{author}{\bibfnamefont{D.~M.} \bibnamefont{Greenberger}},
  \bibinfo{author}{\bibfnamefont{M.~A.} \bibnamefont{Horne}},
  \bibinfo{author}{\bibfnamefont{A.}~\bibnamefont{Shimony}}, \bibnamefont{and}
  \bibinfo{author}{\bibfnamefont{A.}~\bibnamefont{Zeilinger}},
  \bibinfo{journal}{Am. J. Phys} \textbf{\bibinfo{volume}{58}},
  \bibinfo{pages}{1131} (\bibinfo{year}{1990}).

\bibitem[{\citenamefont{Papaliolios}(1967)}]{pap}
\bibinfo{author}{\bibfnamefont{C.}~\bibnamefont{Papaliolios}},
  \bibinfo{journal}{Physical Review Letters} \textbf{\bibinfo{volume}{18}},
  \bibinfo{pages}{622} (\bibinfo{year}{1967}).

\end{thebibliography}
\end{document}